\documentclass{agujournal2019}

\usepackage{url} 

\journalname{JGR-Space Physics}

\usepackage{graphicx,color}
\usepackage{epstopdf}
\usepackage{amsmath}
\usepackage{amsfonts,amsbsy,amssymb,xspace}
\usepackage{layout}
\usepackage{epsf}
\usepackage{times}
\usepackage[section]{placeins}

\begin{document}

%
%


\title{A drift-kinetic method for obtaining gradients in plasma properties from single-point distribution function data}

%
%




\authors{B. A. Wetherton\affil{1}, J. Egedal\affil{1}, P. Montag\affil{1}, A. L\^e\affil{2}, and W. Daughton\affil{2}}


\affiliation{1}{Department of Physics, University of Wisconsin-Madison, Madison, Wisconsin 53706, USA}
\affiliation{2}{Los Alamos National Laboratory, Los Alamos, New Mexico 87545, USA}






\correspondingauthor{Blake Wetherton}{bwetherton@wisc.edu}


 \begin{keypoints}
 \item	A drift-kinetic method of measuring spatial gradients in the distribution function from local velocity-space distributions is developed.
 \item This framework is extended to find gradients in arbitrary moments of the distribution function. 
 \item These methods are verified with simulation data, demonstrating resolution at the electron kinetic scale.
 
 \end{keypoints}


%

%
%


\begin{abstract}
In this paper, we derive a new drift-kinetic method for estimating gradients in the plasma properties through a velocity space distribution at a single point. The gradients are intrinsically related to agyrotropic features of the distribution function. This method predicts the gradients in the magnetized distribution function, and can predict gradients of arbitrary moments of the gyrotropic background distribution function. The method allows for estimates on density and pressure gradients on the scale of a Larmor radius, proving to resolve smaller scales than any method currently available to spacecraft. The model is verified with a set of fully-kinetic VPIC particle-in-cell simulations.
\end{abstract}

%
%

%


%
%
%
%


\section{Introduction} \label{intro}
The properties of the plasma in the Earth's magnetosphere as well as the connection between large scale plasma dynamics on the Sun and the near Earth environment have  been studied intensely over the past decades using increasingly sophisticated spacecraft. With a few exceptions, most of these studies have been carried out through the use of a single spacecraft. A significant and fundamental problem to interpreting spacecraft measurements is that the relative speed between large-scale magnetic structures and the spacecraft is generally not known. This often makes it impossible to characterize the length scales of the dynamical plasma structures encountered because it is not possible to distinguish time variation from spatial variation. The problem can be overcome by applying several spacecraft flying in close formation \cite{Dunlop1988,Chanteur1998}; length scales on the order of the spacecraft separation can then be determined accurately.

The resolution of fine scales is crucial to the understanding of many processes in collisionless plasma physics. Of particular interest, in magnetic reconnection the thickness of the current layer can be on the electron kinetic scale, while various terms in the generalized Ohm's law, which can be written as
\begin{equation}
\label{eq:Ohmslaw} {\bf E}+{\bf v} \times {\bf B}=\eta {\bf j} +
\frac{1}{ne}\left({\bf j}\times {\bf B}- \nabla\cdot{\bf p}_e\right)  +
\frac{m_e}{n e^2} \frac{d{\bf j}}{dt} \quad,
\end{equation}
decouple at different scales; for example, $\boldsymbol \nabla \cdot \bf p_e$ becomes important for gradient scales on the order of the thermal electron Larmor radius $\rho_{e} = m_e v_{th}/e B$ or electron skin depth $d_e = c/\omega_{pe}$. Thus, to characterize the terms down to the kinetic scale would require a tight spacecraft formation to fully resolve, which will also sacrifice overall coverage. Determining local gradients go a long way towards interpreting the overall structure of a current layer. 

In this paper, we develop new methods that allow length scales of plasma structures to be determined at spatial scales as small as the electron Larmor radius $\rho_{e} \sim \sqrt{\beta_e} d_e$,  allowing for $d_e$ scale gradients to be accurately characterized if $\beta_e$ is not too large. Our methods could be implemented using the full three dimensional electron distribution function, which is now available with sufficiently fast time resolution from the Fast Plasma Investigation (FPI) instrumentation suite of NASA's Magnetospheric Multiscale (MMS) mission \cite{Pollock2016}, and will likely be available to future spacecraft missions. The methods make a connection between the apparent agyrotropy of magnetized distributions and gradients in plasma properties perpendicular to the magnetic field lines.

Agyrotropy is the breaking of the symmetry of a distribution function about the magnetic field line, and it is commonly used as a signature for the demagnetization of electrons near the x-line in magnetic reconnection \cite{Scudder2008}. In principle, in a well-magnetized plasma, the fast motion of the gyroorbit will cause the distribution to be constant about its nearly circular trajectory; therefore, a departure from gyrotropy is often implied to be the result of the demagnetization of the particle species. For the electrons in reconnection, this would happen in the electron diffusion region. Several measures of agyrotropy (sometimes called nongyrotropy) have been developed \cite{Scudder2008,Aunai2013,Swisdak2016}, generally measuring the deviation from a diagonal pressure tensor with entries ($p_{||},p_\perp,p_\perp$) in a magnetic field aligned basis. In these measures, agyrotropy is strong not only in the electron diffusion region, but also along the topological boundaries formed by the separatrices. Thus, in these measures, agyrotropy is not a unique signature of the electron diffusion region. The agyrotropy associated with the separatrices is based on a transition between two topological regions on the scale of a Larmor radius. While electron distributions measured at the separatrix are strongly agyrotropic in the frame of the reconnection region, taking the drift-kinetic approach of measuring $f(\mathbf x_{gc})$ to be the phase space density of particles with guiding centers at $\mathbf x_{gc} = \mathbf x - \boldsymbol \rho(\mathbf v)$ rather than current position at $\mathbf x$ can in many cases lead to gyrotropic  distributions.  This tells the story of separate well-magnetized plasma populations interpenetrating at the depth of a Larmor radius, sometimes resulting in crescent-type distributions \cite{Egedal2016}.

Recent observational work on magnetic reconnection has emphasized these agyrotropic crescent distributions. For example, Burch et al. \cite{Burch2016} found the presence of crescent-shaped distributions, both in the perpendicular plane and a parallel-perpendicular plane. This sort of agyrotropic distribution is thought to be a hallmark of the electron diffusion region. However, as noted above, crescent shaped distributions can also a result of crossing the separatrix, where large density gradients exist in asymmetric reconnection. These highly-agyrotropic distributions can be seen as hallmarks of strong gradients in the reconnection geometry. In this paper, we will explicitly link the agyrotropy of the electron distribution function with spatial gradients. Section \ref{sec:agyro} contains an analysis of the effects of density gradients on the commonly-used measures of agyrotropy. In Section \ref{deriv}, we build up a framework to characterize length scales smaller than the separation distance between spacecraft, then verify it using data from several VPIC fully-kinetic simulations in Section \ref{PIC}, demonstrating that this method can accurately characterize gradients in density and pressure on the scale of $\rho_{e}$. 
The paper concludes with a discussion in Section \ref{conc}.

\section{Apparent agyrotropy of simple distributions that are gyrotropic in the guiding center frame} \label{sec:agyro}

As mentioned in Section \ref{intro}, strongly agyrotropic distributions are often associated with regions where plasma properties vary sharply. Previous work has also noted the relationship between agyrotropy and gradients at the scale of the Larmor radius \cite{Scudder2012,Scudder2015}, but have focused primarily on reconnection scenarios rather than simple model equilibria with density gradients. In this section, we compute measures of agyrotropy for a simple guiding center distribution with a spatially varying density. We choose the simplest magnetized guiding center distribution with perpendicular density gradients.

\begin{equation} \label{eq:gcdist}
    \bar f = \left( \frac{m}{2 \pi T} \right)^\frac{3}{2} \left(n_0 + x \nabla n\right) e^{-\frac{m |\mathbf v|^2}{2T}}
\end{equation}

We choose $\mathbf B = B \hat{z}$ and $T$ to be constant for simplicity. We will not have an electric field in this example, but it can be shown that an arbitrary perpendicular electric field will not change the results of the calculation of the agyrotropy parameters, though the intermediate steps will be more complicated and include a velocity shift to the $\mathbf E \times \mathbf B$ frame. An electric field of $\mathbf E = T \nabla n/(n q B) \hat x$ is important, as it will allow the species to be in fluid and drift-kinetic equilibrium. We note that the distribution of Eq.~2 is binned by the location of the guiding center of the particle rather than the particle's instantaneous position, as a spacecraft will typically measure. As such, to evaluate the local distribution function $f(\mathbf x, \mathbf v)$, we must account for the shift of the vector Larmor radius $\boldsymbol \rho$. We now calculate 

\begin{equation}
    f = \bar f \left( \mathbf x - \boldsymbol \rho \right) = \left( \frac{m}{2 \pi T} \right)^\frac{3}{2} \left(n_0 + \nabla n \left(x - \frac{m v_y}{q B}\right) \right) e^{-\frac{m |\mathbf v|^2}{2T}} \quad .
\end{equation}
Notably, in this toy model $f$ becomes negative for particles with large enough $v_y$. This means that particles with larger Larmor radii than the gradient scale will be represented by an unphysical phase space density due to the simplified form of Eq.~2, but this is a negligible contribution for gradient scales that are small relative to the thermal Larmor radius. Having an analytic form of the distribution function allows us to take moments at $x=0$. It can easily be seen that the density moment gives us $n = n_0$. The bulk velocity moment reflects the diamagnetic drift. 
\begin{equation}
\mathbf u = -\frac{T \nabla n}{n q B} \hat y
\end{equation}
With this, the pressure tensor can be straightforwardly calculated as
\begin{align}
        \mathbf P  = &  \left( \frac{m}{2 \pi T} \right)^\frac{3}{2} \int \left( \mathbf v - \mathbf u \right) \left( \mathbf v - \mathbf u \right) \left(n_0 - \frac{m v_y}{q B} \nabla n \right) e^{-\frac{m |\mathbf v|^2 }{2T}} \mathrm d^3 v \\
     = & \quad n_0 T \left(\mathbf I - m T\left(\frac{\nabla n}{n_0 q B} \right)^2 \hat y \hat y\right) \quad . \label{eq:ptens}
\end{align}
Simple symmetry arguments show that off-diagonal elements are zero; however, the tensor is still not gyrotropic, as the perpendicular pressures are not identical. If we define $L_\nabla = |n_0/\nabla n|$ and $\rho_{th} = |m v_{th}/ q B|$, we can define $\xi = \rho_{th}/L_\nabla$ as the ratio between the gradient scale and the Larmor scale, and define all of our agyrotropy measures in terms of this parameter. 

We will calculate $A\text{\O}_e/2$ \cite{Scudder2008}, $D_{ng}$ \cite{Aunai2013}, and $\sqrt{Q}$ \cite{Swisdak2016}. All of these parameters measure agyrotropy associated with this gradient, and gyrotropic distributions will return a value of 0. $A\text{\O}_e/2$ and $\sqrt{Q}$ take the value of 1 for a maximally agyrotropic distribution, though $D_{ng}$ has a different normalization that is dependent on $T_{||}/T_\perp$.   

\begin{eqnarray}
  A \text{\O}_e/2 = \frac{\xi^2}{2 + \xi^2}\\
  D_{ng} =  \frac{\sqrt{2} \xi^2}{3+\xi^2}\\
  \sqrt{Q} = \frac{\xi^2}{\sqrt{(2-\xi^2)(6-\xi^2)}}
\end{eqnarray}

\begin{figure}[h!]
	\centering
	\includegraphics[width=1.0\linewidth]{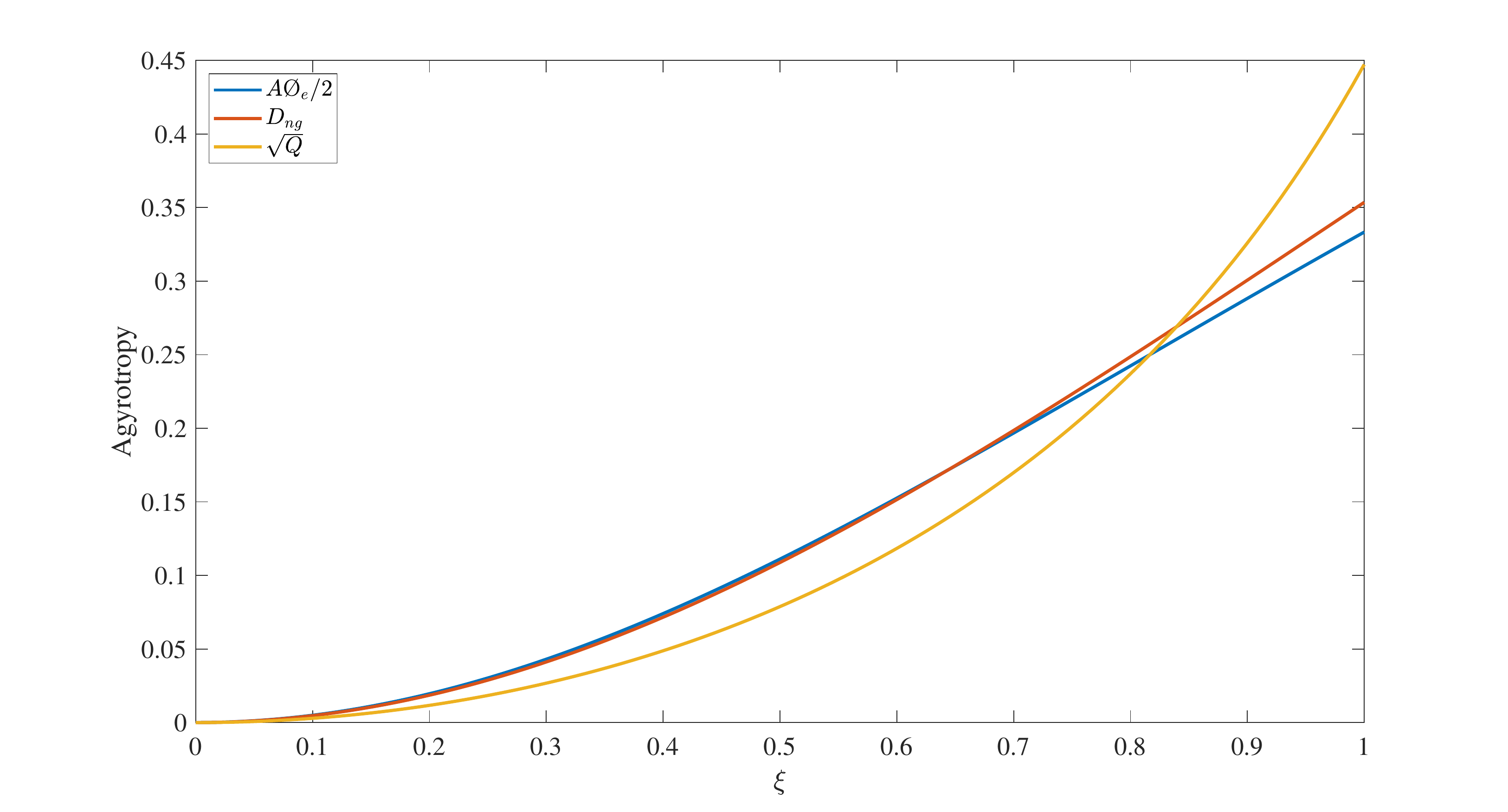}
	\caption{$A\text{\O}_e/2$ and $\sqrt{Q}$ plotted against the ratio of the Larmor scale to the gradient scale, $\xi$, for the relevant range. }
	\label{fig:agyros}
\end{figure}

$\sqrt{Q}$ clearly has issues for $\xi>1$, but that is a result of the nonphysical behavior of this model distribution in that regime (as can easily be seen in Eq.~\ref{eq:ptens}). All three measures reduce to zero in the gradient-free limit. A plot of $A\text{\O}_e/2$ and $\sqrt{Q}$ can be seen in Fig.~ \ref{fig:agyros}. $\sqrt{Q}$ is less sensitive to small gradients than the other two measures, but all three measures are increasing functions of the strength of the density gradient, even though the guiding center distribution is perfectly isotropic. This allows for the possibility of determining unknown gradients through a measure of the agyrotropy of a distribution function. 

\section{Theoretical basis for length scale characterization} \label{deriv}
\begin{figure}[h]
	\includegraphics[width=1.0\linewidth]{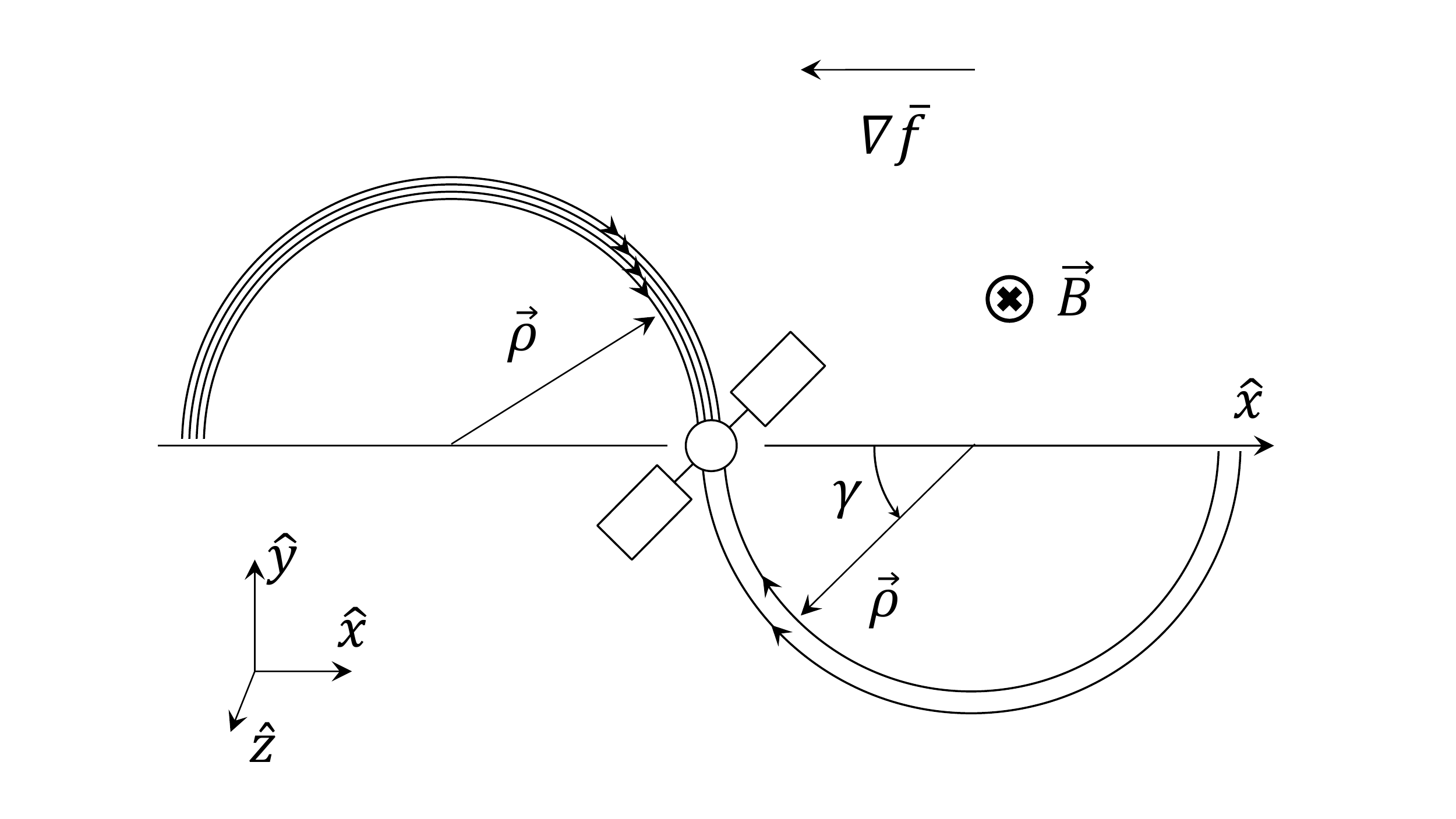}
	\caption{Illustration of how a single spacecraft sampling the electron distribution can be applied for characterizing the gradient in the gyrotropic electron distribution function $\nabla \bar f$. }
	\label{fig:sat}
\end{figure}

In the previous section, we considered a simple example and showed that agyrotropy develops as a result of gradients in the distribution function. In this section, we will consider more general geometries and rigorously show how gradient scales can be inferred through measurements of the distribution function. Before deriving the model in detail, we first provide a heuristic description of how
the plasma length scales can be obtained from  electron
distributions measured by a single spacecraft.
Fig.~\ref{fig:sat} illustrates a model geometry of a spacecraft
sampling the electron distribution $f$. We assume that there is a
gradient in $f$ pointing in the negative $x$ direction. With ${\bf B}$
in the negative $z$-direction it follows that the flux of
electrons observed in the positive $y$-direction  will be enhanced
while the flux received from the negative $y$-direction is
reduced. Furthermore, considering the separation of the respective
guiding centers ($2\rho_{e}$) in Fig.~\ref{fig:sat}, it is clear that
the relative difference in these fluxes must be given by
$2\rho_{e}\nabla f /f$, where $\rho_e$ is the electron Larmor radius for
the energy considered. This anisotropy of particle flux is the origin of the diamagnetic drift.

The approach outlined with the heuristic arguments above is made concrete in this paper.  We can rigorously derive expressions for perpendicular distribution function gradients starting from the kinetic Vlasov equation governing collisionless plasma. Although we are primarily interested in properties of the electron distribution, we will derive expressions for a general species in the drift kinetic limit. By inserting the appropriate mass, charge, and distibution, the electron equations are easily recovered. We begin by noting, as is discussed in great detail in the book by Hazeltine and Meiss \cite{hazeltine:1992}, that in the drift-kinetic limit the Vlasov equation imposes that the first-order expectation of the variation from gyrotropy $\tilde{f}$  can be expressed as:
\begin{equation}
\tilde{f}(\mathbf{x},U,\mu,\gamma,t) = \boldsymbol{\rho} \cdot \left[ q \frac{\partial \mathbf{A}}{\partial t} \frac{\partial \bar{f}}{\partial U} - q \left( \mathbf{b} \times \mathbf{v}_D \right) \frac{\partial \bar{f}}{\partial \mu} - \boldsymbol{\nabla} \bar{f} \right] + \frac{v_\parallel \mu}{\Omega_s} \frac{\partial \bar{f}}{\partial \mu} \left(\hat{\boldsymbol{\rho}}\hat{\mathbf{v}}_\perp : \boldsymbol{\nabla} \mathbf{b} - \frac{1}{2}\mathbf{b}\cdot \boldsymbol{\nabla} \times \mathbf{b} \right) \quad.
\label{fgyro}
\end{equation}
Here $\bar{f}= \bar{f}(\mathbf x,U,\mu, t)$ is the gyro-averaged distribution, $U$ is total particle energy (kinetic plus an electrostatic potential), $\Omega_s = q B/m$ is the signed cyclotron frequency, $\mu$ is the (first adiabatic invariant) magnetic moment, $\gamma$ is the gyrophase such that $\boldsymbol \rho = \mathbf b \times \mathbf v / \Omega_s = \rho (\mathbf e_{\perp 1} \sin \gamma + \mathbf e_{\perp 2} \cos \gamma$), with $(\mathbf b, \mathbf e_{\perp 1}, \mathbf e_{\perp 2})$ forming a right-handed local coordinate system, and
\begin{equation}
\mathbf{v}_D = \frac{\mathbf{E} \times \mathbf{B}}{B^2} + \frac{1}{\Omega_{{s}}} \mathbf{b} \times \left( \frac{\mu}{m} \boldsymbol{\nabla} B + v_\parallel^2 (\mathbf{b}\cdot \boldsymbol{\nabla}) \mathbf{b} + v_\parallel \frac{\partial \mathbf{b}}{\partial t} \right)
\end{equation}
is the drift velocity expected for a particle at each location in phase space. We are most interested in the gradient information that can be recovered from an individual spacecraft that bins the distribution as a function of velocity rather than the adiabatic invariants. Thus, we note that the change in coordinates to $(v_\parallel,v_\perp)$ space provides a mixing between coordinate and velocity spaces,

\begin{equation}
\boldsymbol{\nabla}_{U,\mu} = \boldsymbol{\nabla}_{v_\parallel,v_\perp} - \boldsymbol{\nabla} \mu \frac{\partial}{\partial \mu} - \boldsymbol{\nabla} U \frac{\partial}{\partial U} \quad .
\end{equation}
Importantly, this eliminates any contribution to $\tilde f$ for the $\boldsymbol \nabla B$ drift and alters the inductive electric field in Eq.~\ref{fgyro} to be the full electric field. 

By multiplying each side of Eq.~\ref{fgyro} by the vector Larmor radius $\boldsymbol{\rho}$ and integrating over the gyrophase $\gamma$, we find the perpendicular component of the gradient of $\bar{f}$:
\begin{equation}
\boldsymbol{\nabla}_{\! \!\perp} \bar{f} = - q \mathbf{E}_\perp \frac{\partial \bar{f}}{\partial \mathcal{E}} - q \left( \mathbf{b} \times \mathbf{v}_D \right) \frac{\partial \bar{f}}{\partial \mu}-\frac{1}{\pi \rho^2} \int_0^{2\pi} \boldsymbol{\rho} f \mathrm{d} \gamma
\label{gradf}
\end{equation}
Notably, the terms not dotted into the gyroradius in Eq.~\ref{fgyro} integrate out to zero. It can also be shown that the $\mathbf{E} \times \mathbf{B}$ drift term in Eq.~\ref{fgyro} combines with the $\mathbf{E}_\perp$ term to form $e \mathbf{E}_\perp \partial{\bar{f}}/\partial{\mathcal{E}_{\! \perp}}$ in an $(\mathcal{E}_{\! \perp},\mathcal{E}_{ \parallel})$ basis:
\begin{equation}
\boldsymbol{\nabla}_{\! \!\perp} \bar{f} = - q \mathbf{E}_\perp \frac{\partial \bar{f}}{\partial \mathcal{E}_\perp} + \left(m v_{||}^2 (\mathbf b \cdot \boldsymbol \nabla)\mathbf b  + m v_{||} \frac{\partial \mathbf b}{\partial t}\right) \left(\frac{\partial \bar{f}}{\partial \mathcal E_\perp}-\frac{\partial \bar{f}}{\partial \mathcal E_{||}}\right)-\frac{1}{\pi \rho^2} \int_0^{2\pi} \boldsymbol{\rho} f \mathrm{d} \gamma.
\label{gradf2}
\end{equation}
While this expression fully describes $\boldsymbol{\nabla}_{\! \!\perp} \bar{f}$, this is not generally a quantity that is useful to compare to, and the velocity space derivatives must be evaluated judiciously on a spacecraft such as MMS, where there are a finite number of logarithmically-binned energies to evaluate derivatives on. If the goal is still to estimate $\boldsymbol{\nabla}_{\! \!\perp} \bar{f}$, we provide an estimate based on the assumption of a drifting two temperature Maxwellian distribution:

\begin{equation}
\boldsymbol{\nabla}_{\! \!\perp} \ln \bar{f} = \frac{q \mathbf{E}_\perp}{T_\perp} +  \frac{T_\perp-T_\parallel}{T_\perp T_\parallel} \left( 2 \mathcal{E}_\parallel \left[(\mathbf{b}\cdot \boldsymbol{\nabla}) \mathbf{b}\right] + m v_\parallel \frac{ \partial \mathbf{b}}{ \partial t} \right) - \frac{ \int_0^{2\pi} \boldsymbol{\rho} f \mathrm{d} \gamma }{\pi \rho^2 \bar{f}}
\end{equation}

This form can make a justification for dropping all drift terms except $q \mathbf{E}_\perp/T_\perp$ (corresponding to the $\mathbf{E} \times \mathbf{B}$ drift) on the basis of dependence on temperature anisotropy, which will generally be small (as will be the magnitude of the associated drifts). This can be of importance to spacecraft data, as direct measurement of these factors is not generally possible at the single-spacecraft level. While this form contains an abundance of information, it lacks a strong basis for comparison with our intuition on the fluid description of plasma. To this end, we can now take moments of $\boldsymbol{\nabla}_{\! \!\perp} \bar{f}$ to find the perpendicular gradients of fluid quantities. In particular, we define 

\begin{equation} \label{momdef}
\mathcal{M}_{k,l} = \int v_\perp^k v_\parallel^l \bar{f} \mathrm{d}^3 v = \int v_\perp^{k+1} v_\parallel^l \bar{f} \mathrm{d} v_\parallel \mathrm{d} v_\perp \mathrm{d} \gamma \quad .
\end{equation}

It can be shown that the $\boldsymbol{\nabla}_{\! \! \perp}$ operator commutes with the moment operator on $\bar{f}$, and thus $\boldsymbol{\nabla}_{\! \! \perp} \mathcal{M}_{k,l}$ = $\int v_\perp^k v_\parallel^l \boldsymbol{\nabla}_{\! \! \perp} \bar{f} \mathrm{d}^3 v$. Evaluating these integrals, we find:
\begin{multline}
\boldsymbol{\nabla}_{\! \! \perp} \mathcal{M}_{k, l} = -  2 \Omega_s \mathbf b \times \int \mathbf{v}_\perp v_\parallel^l v_\perp^{k-2} f \mathrm{d}^3 v +
\frac{2 \pi q \mathbf{E}_\perp}{m} \left[ \delta_{k 0} \int v_\parallel^l \bar{f}_\parallel \mathrm{d} v_\parallel + k \int v_\parallel^l v_\perp^{k-1} \bar{f} \mathrm{d} v_\perp \mathrm{d} v_\parallel \right] \\
+  (\mathbf{b} \cdot \boldsymbol{\nabla}) \mathbf{b} \left[ (l + 1) \mathcal{M}_{k, l} - 2 \pi \left( \delta_{k 0} \int v_\parallel^{l+2} \bar{f}_\parallel \mathrm{d} v_\parallel + k \int v_\parallel^{l+2} v_\perp^{k-1} \bar{f} \mathrm{d} v_\perp \mathrm{d} v_\parallel  \right) \right]\\
+  \frac{\partial \mathbf{b}}{\partial t} \left[ l \mathcal{M}_{k, l-1} - 2 \pi \left( \delta_{k 0} \int v_\parallel^{l+1} \bar{f}_\parallel \mathrm{d} v_\parallel + k \int v_\parallel^{l+1} v_\perp^{k-1} \bar{f} \mathrm{d} v_\perp \mathrm{d} v_\parallel \right) \right] \quad ,
\end{multline}
where $\delta_{k0}$ is the Kronecker delta and we have used the shorthand $\bar{f}_\parallel$ to identify $\bar{f}(v_\perp = 0)$. We note that the drift terms not including $\bar{f}_\parallel$ have resulted from integration by parts in $v_\perp$ in Eq.~\ref{gradf2}. The $\bar{f}_\parallel$ terms no longer have $v_\perp$ coefficients, and are thus integrals of a full derivative in $v_\perp$, resulting in the values at the bounds of the integral. If we define 
\begin{eqnarray}
\mathcal{M}_{-2,l} = 2 \pi \int v_\parallel^l \bar{f}_\parallel \mathrm{d} v_\parallel\\
\mathcal{M}_{-1,l} = 2 \pi \int v_\parallel^l \bar{f} \mathrm{d} v_\parallel \mathrm{d} v_\perp ,
\end{eqnarray}
where $\mathcal{M}_{-1,l}$ follows the definition of Eq.~\ref{momdef}, but $\mathcal{M}_{-2,l}$ does not (as a result of the aforementioned integration of a full derivative), this can be written slightly more concisely as:
    \begin{multline}
    \label{eq:moments}
 \boldsymbol{\nabla}_{\! \! \perp} \mathcal{M}_{k l} = -  2 \Omega_s \mathbf b \times \int \mathbf{v}_\perp v_\parallel^l v_\perp^{k-2} f \mathrm{d}^3 v +
 \frac{q \mathbf{E}_\perp}{m}  (k+ \delta_{k 0}) \mathcal{M}_{k-2, l}\\
 +  (\mathbf{b} \cdot \boldsymbol{\nabla}) \mathbf{b} \left[ (l + 1) \mathcal{M}_{k l} - (k+ \delta_{k 0}) \mathcal{M}_{k-2, l+2}  \right] \\
 +  \frac{\partial \mathbf{b}}{\partial t} \left[ l \mathcal{M}_{k ,l-1} - (k+ \delta_{k 0}) \mathcal{M}_{k-2, l+1} \right]  \quad .
 \end{multline}

Of particular interest are gradients in density and pressure. In this gyrotropic definition of $\bar{f}$, we note that 
\begin{eqnarray}
n = \mathcal{M}_{0,0}\\
p_\perp = \frac{m}{2} \mathcal{M}_{2,0}
\end{eqnarray}
By evaluating Eq. \ref{eq:moments}, we then find that
\begin{multline}
\label{gradn}
\boldsymbol{\nabla}_{\! \! \perp} n = \frac{2 \pi q \mathbf{E}_\perp}{m} \int \bar{f}_\parallel \mathrm{d} v_\parallel + (\mathbf{b} \cdot \boldsymbol{\nabla}) \mathbf{b} \left( n - 2 \pi \int v_\parallel^2 \bar{f}_\parallel \mathrm{d} v_\parallel \right) \\- \frac{\partial \mathbf{b}}{\partial t} \left( 2 \pi \int v_\parallel \bar{f}_\parallel \mathrm{d} v_\parallel \right) - 2 \int \frac{\boldsymbol{\rho} f}{\rho^2} \mathrm{d}^3 v
\end{multline}

\begin{equation}
\label{gradpperp}
\boldsymbol{\nabla}_{\! \! \perp} p_\perp = n q \left( \mathbf{E}_\perp + \mathbf{u}_s \times \mathbf{B} \right) - m n u_\parallel \frac{\partial \mathbf{b}}{\partial t} + (\mathbf{b} \cdot \boldsymbol{\nabla}) \mathbf{b} \left(p_\perp-p_\parallel -m n u_\parallel^2 \right) .
\end{equation}

These equations present the best estimate of the gradients, but several terms are not locally measured by a single spacecraft. We can see that in the absence of the $\partial \mathbf{b}/\partial t$ and $(\mathbf{b} \cdot \boldsymbol{\nabla}) \mathbf{b}$ terms (which are not readily available to single spacecraft observation), the perpendicular pressure gradient term becomes equivalent to the statement that the nonideal electric field is entirely attributable to the diamagnetic drift. This is somewhat unfortunate, as it means that using only the measurements readily available to a single spacecraft, this model does not allow for Ohm's Law terms that can account for reconnection. As such, in this framework, not much can be learned from a single spacecraft about non-ideal dynamics within electron diffusion regions. However, if one desires to estimate pressure gradients for reasons other than determining the breaking of the frozen-in law, this method should provide a good estimate for most magnetized plasma environments that do not simultaneously experience strong curvature and temperature anisotropy. Notably, anisotropy is measurable at the single-spacecraft level, and strong curvature will be accompanied by sharp shifts in the time series measurement of $\mathbf b$, so regions where dropping the curvature term may cause significant error should be identifiable. We also note that it may be possible to estimate $\partial \mathbf{b}/\partial t$ and $(\mathbf{b} \cdot \boldsymbol{\nabla}) \mathbf{b}$ by matching their contributions to Eq.~\ref{gradpperp} to the remaining terms in Eq.~\ref{eq:Ohmslaw}, or a multi-spacecraft reconstruction of the local magnetic geometry.

\section{PIC verification of length scale characterization} \label{PIC}
In order to verify the drift-kinetic model's ability to characterize density and pressure gradients in a plasma, we calculate the gradients derived from the model on particle data obtained from a series of fully-kinetic VPIC \cite{bowers:2009} particle-in-cell simulations. While these simulations are two dimensional, mathematically our methods should apply equally well for fully three dimensional reconnection scenarios. The simulations are performed in a modified Harris sheet configuration \cite{roytershteyn:2012} at ${m_i}/{m_e} = 400$ with a variety of background density asymmetries representing the low density magnetosphere and high density magnetosheath (herein denoted as ${n_1}/{n_2}$). The runs correspond to the setup of a suite of simulations described in Chapter 3 of \cite{MontagThesis} and are antiparallel. This means that the electrons will not be magnetized everywhere, but this is a small region of the simulations, and demagnetized electrons are soon remagnetized. As such, the results of Section \ref{deriv} should hold over most of the simulation domain. These simulations use periodic boundary conditions in the $x$-direction and conducting/reflecting boundaries in the $z$-direction, have a domain size of $6720 \times 3360$ cells $= 80 \hspace{2 pt} d_i \times 40 \hspace{2 pt} d_i$ based on the higher upstream density $n_1$, and employ the reduced value of $\omega_{pe}/\omega_{ce} = 1.5$, with $\beta_1 = 3$. In total, each run contains $\sim 18$ billion numerical particles. 

We evaluate gradients in two ways: via an implementation of the drift-kinetic method (Eqs.~\ref{gradn} and \ref{gradpperp}) directly on particle data and via direct finite difference calculation of spatial gradients of the density and pressure profiles. The PIC distribution functions are created from particles within a box of approximately $2 \hspace{2 pt} d_e \times 6 \hspace{2 pt} d_e$ (containing on average $\sim 80,000$ electron particles), with a measurement centered every $1 \hspace{2 pt} d_e$ in the $N$ direction. We note that there is still a considerable amount of noise in our density gradient measurements at this domain size, but we do some smoothing to present the picture of the full domain. We present the data in a normalized form that represents the inverse gradient length scale in terms of the electron inertial length to indicate the fine scale structure encountered in the reconnection geometry.  

\begin{figure}[h!]
	\centering
	\includegraphics[width=1.0\linewidth]{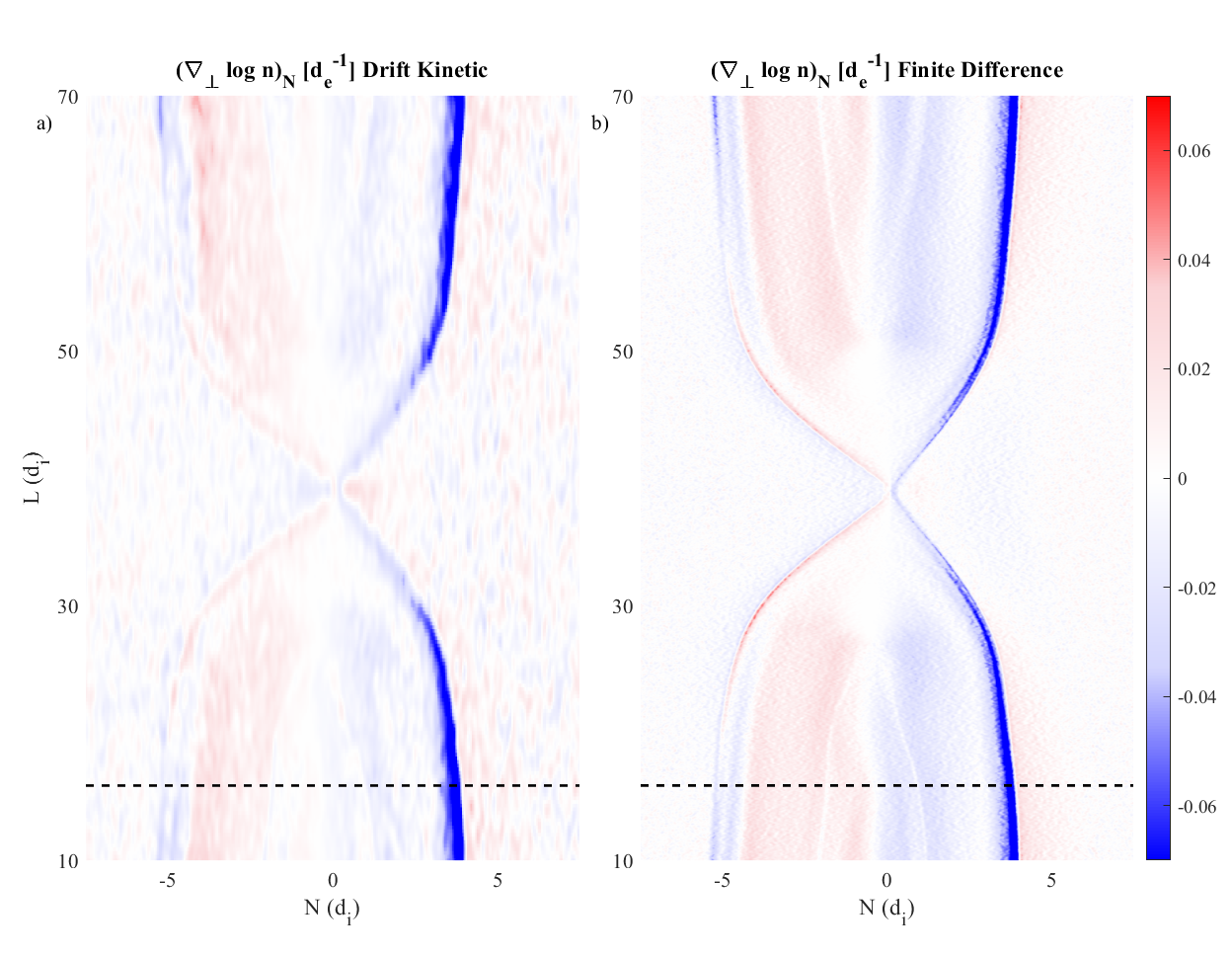}
	\caption{Normalized components of $\boldsymbol{\nabla}_{\! \! \perp} \log n$ for a simulation of antiparallel reconnection with ${n_1}/{n_2} = 1.4$ through Eq.\ref{gradn} and a direct finite difference method. The dashed line represents the cut taken in Figure \ref{fig:asymfig}a). }
	\label{fig:ncomp14}
\end{figure}

\begin{figure}[h!]
	\centering
	\includegraphics[width=1.0\linewidth]{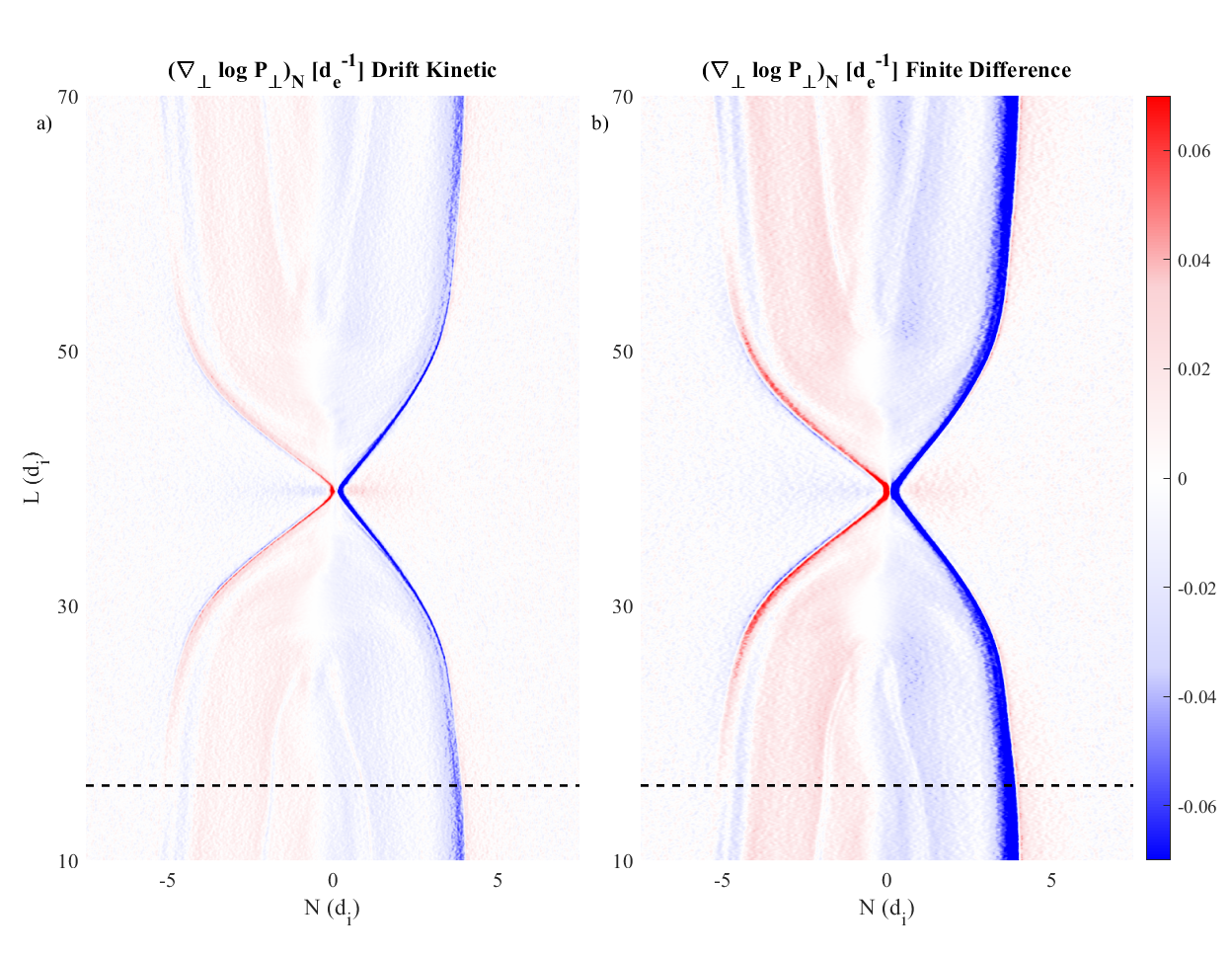}
	\caption{Normalized components of $\boldsymbol{\nabla}_{\! \! \perp} \log p_\perp$ for a simulation of antiparallel reconnection with ${n_1}/{n_2} = 1.4$ through Eq.\ref{gradpperp} and a direct finite difference method. The color scale is slightly saturated to emphasize the quality of the low amplitude match. The dashed line represents the cut taken in Fig.~\ref{fig:asymfig}a). }
	\label{fig:pperpcomp14}
\end{figure}

\begin{figure}[h!]
	\centering
	\includegraphics[width=1.0\linewidth]{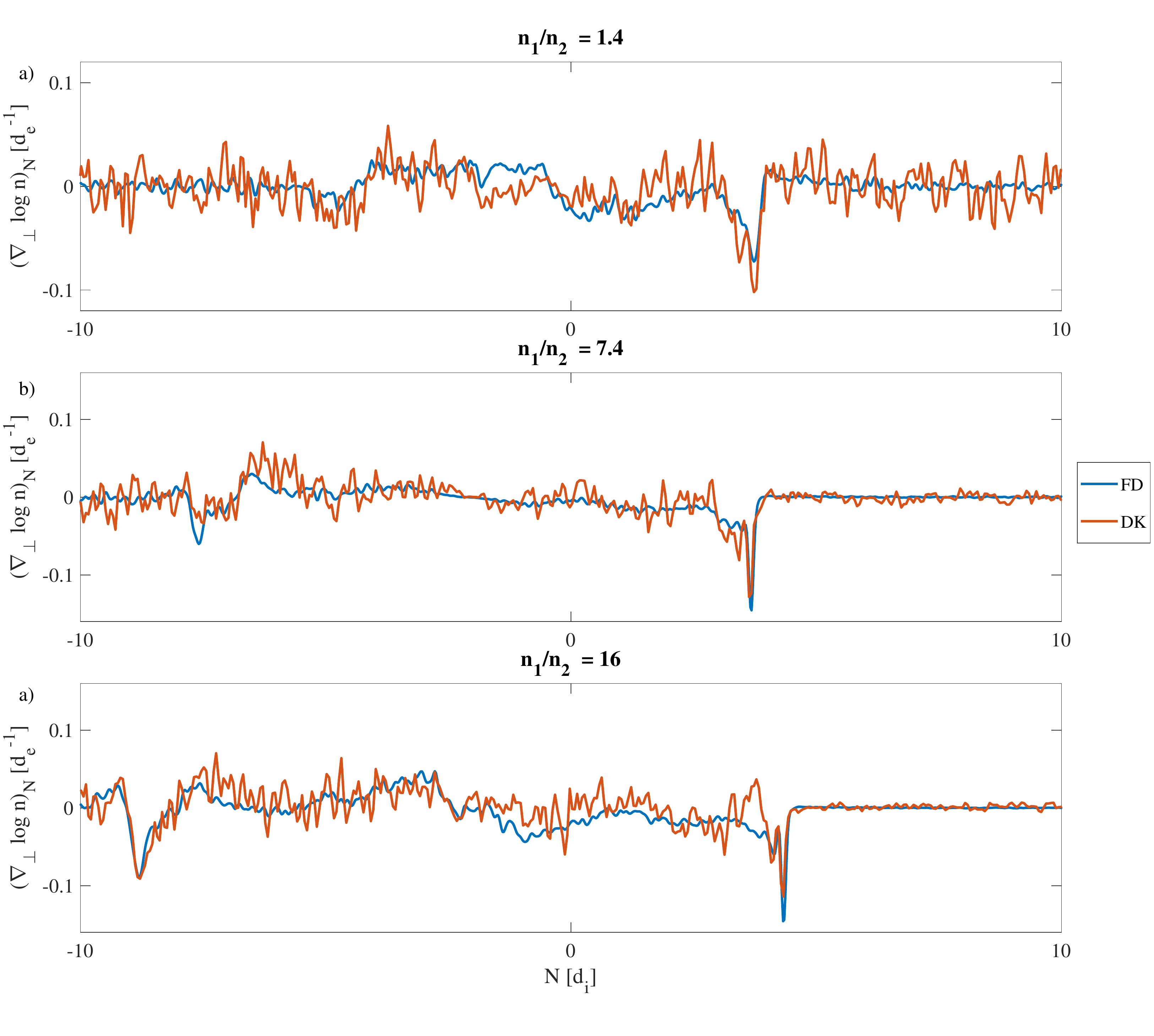}
	\caption{Cuts of normalized $N$ components of $\boldsymbol{\nabla}_{\! \! \perp} \log n$ for simulations of antiparallel reconnection with ${n_1}/{n_2} = 1.4, 7.8,$ and $16$ through Eq.\ref{gradn} and a direct finite difference method. }
	\label{fig:asymfig}
\end{figure}

\FloatBarrier
It should be noted that there are some limits to the validity of this model. Most importantly, in strongly asymmetric reconnection with low guide field, strong electric fields tend to develop with a width on the order of the electron Larmor radius on the low density separatrix. In these conditions, the drift kinetic assumptions are violated, and our framework significantly overestimates the gradients. The model should be valid if $\boldsymbol{\rho}_e \cdot \boldsymbol{\nabla} v_{EB} \ll v_{the}$, and if gradient scales are larger than the electron Larmor radius. The fields at the separatrix in strongly asymmetric reconnection produce $v_{EB} \sim v_{the}$ with a width on the order of a few electron Larmor radii, and this is not easily overcome in this model framework, but elsewhere, the model assumptions are generally satisfied. Figure \ref{fig:asymfig} shows a comparison of $\boldsymbol{\nabla}_{\! \! \perp} n$ calculated by the two methods for a variety of upstream density asymmetry values in antiparallel reconnection. For the antiparallel run with density asymmetry of 16 presented, our model is observed to provide an accurate estimate of the density gradient at the separatrices, and accuracy is expected only to improve for configurations including a guide magnetic field.

We do note that these simulations do not fully encompass the parameters of magnetospheric reconnection. In particular, $\beta_1 = 3.0$ is rather high, leading to weaker electric fields at the separatrix and an effectively higher thermal speed. This helps to keep the model within its limits. In realistic magnetopause conditions, a range of upstream $\beta$, as well as temperature and density asymmetries must be considered. Testing the model in a separate simulation designed to match the event of \cite{Burch2016} (the simulation used in \cite{Egedal2016}) that has lower $\beta_1$ that includes temperature asymmetry, the drift-kinetic model overestimates the gradients at the separatrices by a factor of ~2-3.  Combined with the results of the simulations shown, this implies that the method should be viable for the majority of magnetospheric conditions, though gradient scales can only be trusted to an order of magnitude in cases with some combination of lower $\beta$ and more intense asymmetry than those in the simulations employed in this paper.  

\section{Discussion and Conclusion} \label{conc}

In this paper, we have derived and demonstrated a novel method for inferring plasma gradients from the distribution function measurements of a single spacecraft, linking variations measured within a gyro orbit in velocity space with spatial gradients of a well-magnetized distribution function. This model successfully replicates pressure and density gradients in PIC simulations, and can in principle be applied to MMS data. The PIC verification shows that the gradient estimates can be quite noisy, though they clearly approximate the correct gradients. This may be a challenge when implementing the technique on spacecraft data. Furthermore, it should be noted that this technique requires the electrons to stay well-magnetized to be accurate. As such, in the absence of a strong guide field, this technique is likely not useful immediately at the x-line, though it can be useful in determining the reconnection geometry away from the x-line. 

This model also has implications for the way we think of agyrotropy in distribution functions. Using the drift-kinetic method, we have shown that apparent agyrotropy in the electron distribution function can correspond to spatial gradients in a distribution function that is gyrotropic when spatially sorted to match guiding centers. In this sense, a well-magnetized distribution can be agyrotropic in the standard models of agyrotropy \cite{Scudder2008,Swisdak2016}. Agyrotropy is often used as a measure of demagnetization, but without accounting for spatial gradients in the plasma properties, this is not inherently true. In particular, strong gradients often exist around the separatrices in asymmetric reconnection. While the agyrotropy measure may be high at the separatrix, this does not generally imply that the electrons are demagnetized.

\appendix
\section{Intuitive Derivation of Gradients}
While the derivation in the main body of the paper is the correct one, a more intuitive derivation can reproduce similar results. First, we define a gyrotropic distribution to be one with the property
\begin{equation}
    f(\mathbf{x},\mathbf v, t) = \bar f (\mathbf x - \boldsymbol \rho, \mathbf v - \mathbf{v_D},t).
\end{equation}

Then, we approximate the full distribution function by a first-order Taylor expansion of $\bar{f}$ in position and velocity space. 

\begin{equation} \label{taylor}
    f(\mathbf{x},\mathbf v, t) \approx \bar f(\mathbf{x},\mathbf v, t) - \boldsymbol \rho \cdot \boldsymbol \nabla \bar f \bigg\rvert_{\mathbf x, \mathbf v, t} - \mathbf{v_D} \cdot \frac{\partial \bar f}{\partial \mathbf v} \bigg\rvert_{\mathbf x, \mathbf v, t}
\end{equation}

By multiplying Eq.~\ref{taylor} by $\boldsymbol \rho$, integrating over the gyrophase, and solving for the gradient term, we find 
\begin{equation}\label{weakgf}
    \boldsymbol{\nabla}_\perp \bar f = \frac{\mathbf b \times \mathbf{v_D}}{\rho} \frac{\partial \bar f}{\partial v_\perp} - \int \frac{\boldsymbol \rho f}{\pi \rho^2} \mathrm d \gamma
\end{equation}
In the same coordinate system, Eq.~\ref{gradf} would be written as
\begin{equation} \label{fullgf}
    \boldsymbol{\nabla}_\perp \bar f = \frac{e \mathbf E_\perp}{m v_{||}} \frac{\partial \bar f}{\partial v_{||}} + \frac{\mathbf b \times \mathbf{v_D}}{\rho}\left( \frac{\partial \bar f}{\partial v_\perp} - \frac{v_\perp}{v_{||}} \frac{\partial \bar f}{\partial v_{||}} \right) - \int \frac{\boldsymbol \rho f}{\pi \rho^2} \mathrm d \gamma,
\end{equation}
where the $\mathbf E_\perp$ term cancels with the $\mathbf E \times \mathbf B$ drift term's dependence on the parallel velocity, making Eq.~\ref{weakgf} match Eq.~\ref{fullgf} when the only drift is the  $\mathbf E \times \mathbf B$ drift. The difference in the results of the two derivation then lies in the inclusion of the $\boldsymbol \nabla B$ drift in the Taylor method, and the extra term proportional to $\partial \bar f/ \partial v_{||}$ for the $\partial \mathbf b/ \partial t$ and curvature drifts. If we exclude the curvature drift in this analysis (its terms have a singularity), we can approximate gradients of moments of the distribution function as well. The result will differ from Eq.~\ref{eq:moments} by $(l+1)\mathcal M_{k,l} (\mathbf b \cdot \boldsymbol \nabla) \mathbf b + l \mathcal M_{k,l-1} \partial \mathbf b/ \partial t$.

As such, this simplified approach gives a reasonable heuristic understanding of what the drift kinetic equations are doing, but without some insight into the subtleties of the drift kinetic limit.

\acknowledgments

B.A.W. was supported by the Department of Energy's Computational Science Graduate Fellowship (DOE CSGF) under Grant No. DE-FG02-97ER25308. 
B.A.W would like to acknowledge Dr. Li-Jen Chen for a fruitful discussion about extending the model for pressure gradients. VPIC is an open source code, available at \url{github.com/lanl/vpic}. Results can be reproduced by running simulations with the same initial conditions as listed in the paper.  






%
%
%
%
%
%
%
%
%
%

\bibliography{Mendeley.bib}






\end{document}